# Microwave frequency dissemination systems as sensitive and low-cost interferometers for earthquake detection on commercially deployed fiber cables

**Adonis Bogris(1), Christos Simos(2), Iraklis Simos(1), Thomas Nikas(3), Nikolaos S. Melis(4), Konstantinos Lentas(4), Charis Mesaritakis(5), Ioannis Chochliouros(6), Christina Lessi(6)**

1. University of West Attica, Aghiou Spiridonos, 12243, Egaleo, Athens, Greece, Author e-mail address: abogris@uniwa.gr
2. University of Thessaly, Dept. of Physics, Electronics & Photonics Laboratory, 35100 Lamia, Greece
3. Dept. of Informatics and Telecommunications, National and Kapodistrian University of Athens, Athens, GR-15784, Greece
4. National Observatory of Athens, Institute of Geodynamics, Greece
5. University of the Aegean, Dept. of Information and Communication Systems Engineering, Palama 2, Karlovassi, 83200-Samos-Greece
6. Hellenic Telecommunications Organization S.A. (OTE), 1, Pelika & Spartis,, Maroussi, Athens, Greece

**Abstract:** We experimentally demonstrate a microwave frequency dissemination system operating as a sensitive interferometric sensor of seismic waves on commercially deployed fiber networks in Attika, Greece. Efficient detection of seismic waves from distant epicenters (>400km) is presented © 2021 The Authors.

## 1. Introduction

The use of fiber infrastructures for environmental sensing is attracting global interest, since the optical fiber emerges as a free and easily accessible platform exhibiting a large terrestrial coverage. Moreover, optical fiber networks offer the unique advantage of access in submarine areas where other types of sensors cannot be utilized. Latest experiments demonstrate three dominating techniques that transform optical fiber to a very sensitive sensor of mechanical vibrations coming from either seismic waves, city noise, tsunamis, etc. The distributed acoustic sensing (DAS) is the most popular and mature method/system which offers monitoring at a spatial resolution in the order of a few meters and a very high accuracy [1]. The main drawbacks of DAS are that it can be only used in dark fibers – thus no other communication channels are permitted to co-propagate – is very sensitive to reflections caused by non-ideal connections between different fiber segments in installed deployments and can not operate above 30-40 km. Moreover, DAS tools as commercial products are extremely expensive (> 100 k$) rendering their massive use prohibitive. Another technique proposed by Marra et al [2] relies on the use of ultra-stable laser interferometry. This technique offers superb sensitivity thanks to its coherent nature and is compatible to wavelength division multiplexing (WDM) transmission systems, as it only requires the allocation of a specific wavelength within the grid. Still, it relies on the use of lasers with extreme spectral purity, that is linewidths in the order of Hz, since the system operates with high reliability as long as the coherence length of the laser source is much longer than the fiber link under examination. Lately, another interesting technique was presented by Zhan et al. [3] and relies on the detection of polarization variations at a coherent receiver as a result of fiber deformation along the link. This technique is also compatible to WDM transmission, exploits the already installed polarization multiplexed coherent receivers of long-haul transmission links, however in installed fibers where no coherent detection is used, a customized rather complicated receiver based on polarization demultiplexing is required. In this work, we demonstrate a novel and practical technique that relies on the interferometric use of microwave frequency dissemination in optical fibers, which is compatible to WDM transmission and it constitutes a low-cost implementation with the potential of massive deployment in terrestrial and submarine cables. The technique has been evaluated from July to mid October 2021 in a 50 km (2x25 km) long link in the area of northern Attica and it has been successful in recording earthquakes with magnitudes ranging from M=3 to M=6 from epicenters located even 400 km away from its location (city of Maroussi, Greece).

## 2. Principle of Operation

The technique relies on the dissemination of microwave carriers (> 10 GHz) superimposed on optical wavelengths generated by off-the-shelf lasers. The microwave carrier travels along the fiber and returns back to the transmitter side using a second fiber of the same cable – thus a loopback connection is required at the end of the path. In a commercially deployed WDM network, loopback connection can be performed without disturbing co-propagating channels with the use of WDM demultiplexers and mutliplexers. The signal is detected by a fast photodiode and then is compared with the locally generated carrier with the use of a microwave mixer. This interferometric comparison provides the phase difference between the two RF tones which is proportional to the accumulated propagation delay along the link. Every external disturbance on the fiber causes the modulation of propagation delay as a combined result of variations in both refractive index and fiber length. These changes are imprinted on the phase measurements of our system which are given by $\varphi=2\pi f_{RF} n_g L/c$ where $f_{RF}$ is the microwave frequency, $n_g$ is the refractive index of the fiber, $L$ is the fiber length and $c$ is light velocity. Phase measurements can be easily transformed to strain ($\Delta\varphi/\varphi$), to strain rate ($d(\Delta\varphi/\varphi)/dt$) and can be compared to what seismometers (proportional to strain rate) or accelerometers (first derivative of strain rate) measure.

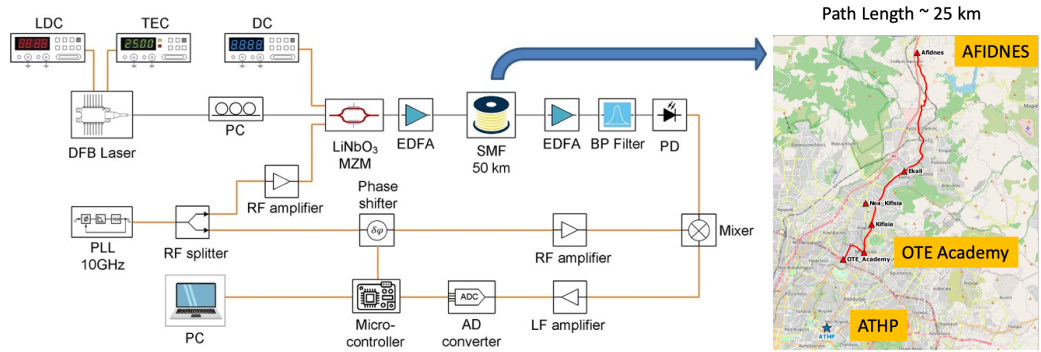

Fig. 1. The Experimental setup of the field trial installed in OTE Academy and a map of the 25 km long fiber path Northern Athens (OTE Academy to Afidnes OTE station)

## 3. Experimental Setup

The experimental setup is sketched in Fig. 1. A polarization-controlled continuous wave DFB laser emitting at 1550nm is modulated by an electro-optical modulator and driven by a 10 GHz phase locked loop frequency generator. The resulting optical signal is amplified with an erbium doped fiber amplifier (EDFA) and launched into a closed loop standard telecommunication optical fiber with a total length of ~52 km and an overall attenuation of ~24 dB. The fiber cable crosses the north part of Attica in Greece and was provided by the Hellenic Telecommunications Organization (OTE). The launched optical power in the fiber is 9 dBm when an EDFA is used. The output signal, after amplification by a second EDFA and optical filtering, is detected by a 10 GHz bandwidth photodiode and the electrical signal is sent to the RF port of a microwave mixer. The LO port of the mixer is driven by a phase-controlled replica of the RF signal that modulates the laser. The phase control is achieved by a microwave phase shifter with a resolution of 6 bit (5.625°). The phase shifter secures that the phase difference between the RF generator and the received signal is around π/2 which leads to the maximum sensitivity. The mixer output is filtered by a low pass filter and then sampled at a rate of 100 Hz with a resolution of 10 bits. The recorded signal is proportional to the overall phase difference $\varphi$ experienced by the optical signal during its propagation in the fiber and therefore carries the signature of thermal variations and mechanical vibrations that stress the fiber. The recorded signal is then digitally processed in order to be converted to strain and reject high frequency noise as well as slow thermal effects. The theoretical sensitivity of the system depends on the optical power that can be injected in the fiber, the RF frequency and the resolution of the analog to digital converter (ADC) at the receiver end. In particular, in the present version of the deployed system, the theoretical sensitivity is limited by the 10-bit resolution ADC and is estimated to be approximately 1.25 mrad, however it could scale down to ~0.3 mrad using a 12-bit ADC. Further improvement could be achieved by increasing the RF modulation frequency and by applying specialized signal processing that could average out the noise. Taking into account the photoelastic effect [4], our implementation that relies on low-cost off-the-shelf components can detect optical path variations in the order of $\Delta L=7\times10^{-8}$ meters. We envisage that this can improved almost two orders of magnitude with the above reported improvements.

## 3. Results and Discussion

The setup described in the previous section has recorded several seismic signals from earthquakes occurring in different regions of Greece during the last three months. Two typical recordings are presented here. The first one (see Fig. 2, left) resulted from a seismic event occurred on July 11, 2021, 00:00:17 UTC, with its epicenter located in the region of Thebes (central Greece) and local magnitude ML=4.3. The epicenter distance from the optical fiber link is approximately 47 km. It is worth noting that this signal was acquired with a relatively low optical signal to noise ratio

(OSNR = 29dB), since neither the EDFA in the transmitter side nor the BP optical filter in the receiver end were used at that time. Subsequent recordings were acquired with an improved OSNR of approximately 43dB. The original raw electronic signal which is proportional to the overall phase difference was first phase unwrapped, then transformed to acceleration (second derivative of $\Delta\varphi$) and finally filtered with a digital low pass filter to reject noise above 5 Hz. For the sake of comparison, we present in the same figure an accelerometric recording of the nearest seismic station (ATHP) operated by the Institute of Geodynamics filtered to the same cutoff frequency. The second recording presented in Fig. 2 (right), resulted from a larger magnitude earthquake that occurred on October 12, 2021, 09:24:03 UTC east of Crete. The distance of the epicenter from the optical fiber is approximately 410 km. Accelerometric recordings of the same nearby seismic station are also presented. Frequency components above 2.5 Hz were rejected. The ATHP broad band accelerometric station is situated 5km SSE of the OTE Academy. It has a 3 component (N-S, E-W and Up-Down) set of force balanced accelerometers. We used one of its horizontal components, taking always into consideration that we observe the transverse seismic wave to the fiber orientation. At both cases P-waves are dumped as we observe the horizontal component and not the vertical (Up-Down), while the S-waves are clearly shown for the same reason. Coda waves can be seen for the Crete earthquake recording due to the distance and magnitude. These are low period signals compared to P and S waves characterizing the tail of the waveform presented in Figure 2. For both earthquakes seismic signals detected with the use of the optical fiber link exhibit high correlation with signals detected from the accelerometric seismic station located nearby, thus proving the efficacy of the proposed system to detect seismic events of different amplitudes and from near and distant locations.

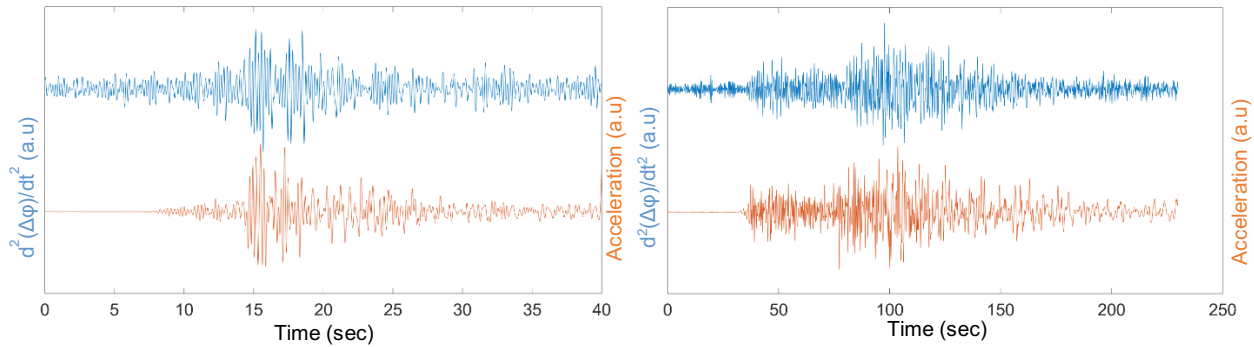

Fig. 2. Comparison between seismic signals recorded with the optical fiber link (blue) and the corresponding signals from a nearby accelerometric station operated by the Institute of Geodynamics, National Observatory of Athens (orange) for two earthquakes in Greece. Left: July 11, 2021, 00:00:17 UTC, Thebes (M=4.3). Right: October 12, 2021, 09:24:03 UTC, Crete (M=6.3)

**Conclusions**

This paper demonstrated an interferometric technique based on microwave frequency dissemination as a sensitive lumped detector of seismicity in commercially deployed fiber networks in Northern Attica, Greece. Next steps will be directed to the extension of the same principle in distributed sensing along the fiber and its comparison with the well-established DAS systems.

**Acknowledgements**

This project has received funding from the University of West Attica. All authors would like to acknowledge Dimitris Polydorou, Diomidis Skalistis and Petros Vouddas from OTE for their efforts in setting up the links that were used in this experiment.